\documentclass[twocolumn,english,prl,singlespace]{revtex4-1}
\usepackage[T1]{fontenc}
\usepackage[latin9]{inputenc}
\usepackage{color}
\usepackage{amsmath}
\usepackage{amssymb}
\usepackage{graphicx}

\makeatletter

\newcommand{\lyxdot}{.}

\@ifundefined{textcolor}{}
{%
 \definecolor{BLACK}{gray}{0}
 \definecolor{WHITE}{gray}{1}
 \definecolor{RED}{rgb}{1,0,0}
 \definecolor{GREEN}{rgb}{0,1,0}
 \definecolor{BLUE}{rgb}{0,0,1}
 \definecolor{CYAN}{cmyk}{1,0,0,0}
 \definecolor{MAGENTA}{cmyk}{0,1,0,0}
 \definecolor{YELLOW}{cmyk}{0,0,1,0}
 }

\@ifundefined{definecolor}
 {\usepackage{color}}{}

\makeatother

\usepackage{babel}
\begin{document}

\title{Classifying directional Gaussian quantum entanglement, EPR steering
and discord}

\author{Q. Y. He$^{1,2*}$, Q. H. Gong$^{1}$ and M. D. Reid$^{2\dagger}$}

\affiliation{$^{1}$State Key Laboratory of Mesoscopic Physics, School of Physics,
Peking University, Collaborative Innovation Center of Quantum Matter,
Beijing, China }

\affiliation{$^{2}$Centre for Quantum Atom Optics, Swinburne University of Technology,
Melbourne, Australia}

\affiliation{$^{*}qiongyihe@pku.edu.cn$, $^{\dagger}mdreid@swin.edu.au$}
\begin{abstract}
Quantum discord quantifies how much Alice's system is disrupted after
a measurement is performed on Bob's. Conceptually, this behavior acts
the same way as quantum steering and we find that the discord grows
with better steering from Bob to Alice. Using Venn diagrams, the relations
between the different classes of Gaussian continuous variable entanglement
and the links to discord for the squeezed-thermal states are established.
We identify a directional quantum teleportation task for each class
of squeezed-thermal state entanglement, and establish a unified signature
for quantum steering, entanglement and discord beyond entanglement.
Quantum steering and discord are promising candidates to quantify
the potential of the directional quantum tasks where Alice and Bob
possess asymmetrically noisy channels. 
\end{abstract}
\maketitle
\textbf{\textcolor{red}{}}The topic of quantum correlations has
received much attention in modern physics \cite{epr,Bell}. Entanglement
is a distinctive feature of quantum correlations \cite{ent} and it
is considered that all entangled states are useful for quantum information
processing (QIP) \cite{ent-qip}. Einstein-Podolsky-Rosen (EPR) correlations
enable error-free predictions for the position $-$ and the momentum
$-$ of one particle given some type of measurement on another. EPR
correlations are especially useful \cite{tele}. As one example, the
fidelity of the quantum teleportation (QT) of a coherent state is
directly related to the strength of EPR correlation available in the
quantum resource \cite{bk}. 

Very recently, there has been an appreciation of the importance of
asymmetry and direction in quantum correlations \cite{steer thmurray,discord,discord vedral,EPRsteering,hw-steering}.
Entanglement is a property shared between two parties, and measures
of it have not been sensitive to differences between the quantum parties
involved \cite{werner-1}. Yet, the original EPR argument was expressed
asymmetrically between the two systems. The analysis by Schrodinger
introduced the asymmetric term {}``steering'' to describe the EPR
idea of one party apparently adjusting the state of another by way
of local measurements \cite{Schrodinger}. This aspect has been beautifully
captured in two recent alternative definitions for quantum correlations:
\emph{quantum discord} \cite{discord,discord vedral} and \emph{EPR
steering} \cite{hw-steering,EPRsteering}. Besides being of intrinsic
fundamental interest, these asymmetrical nonlocalities are attracting
a great deal of attention \cite{Datta,discordnoise,cvexpdiscord,steerexp}
for special tasks in QIP e.g. cloning of correlations \cite{QDcloning},
quantum metrology \cite{metro}, quantum state merging \cite{merging},
remote state preparation \cite{rsp_np_2012} and one-sided device-independent
quantum key distribution \cite{epruses}. Surprisingly, for mixed
states, quantum discord can emerge without entanglement and recent
experiments \cite{expentsep} have used discord to distribute entanglement
using separable states only \cite{expentdis}. Despite the potential
value of directional quantum correlation, relatively little is known
about the quantitative link between discord and steering, and methodologies
to characterise quantum states for their asymmetrical correlation.

Our aim in this Letter is to provide such a characterisation and to
explain the link between discord and steering for the purpose of continuous
variable (CV) QIP. We focus on the subclass of bipartite quantum systems
called Gaussian states \cite{gausdiscord,hw-steering} which have
enabled experimental milestones such as deterministic QT \cite{cvteleexpfuru}.
Asymmetrical Gaussian entanglement and its application to QIP is not
fully understood. To illustrate, it is often interpreted that CV quantum
teleportation (QT) requires a resource with the {}``Duan'' \cite{duan}
symmetric form of entanglement, for which the measures of EPR steering
and discord are largely unaltered if the roles of the two parties
are exchanged \cite{CV tele ,bk,three criteria Buono}. 

Here, we address this gap in knowledge by introducing a classification
of the space of Gaussian states into\textcolor{black}{{} distinct sets
of directional entanglement }classes. We establish the strict relations
between the classes\textcolor{red}{{} }\textcolor{black}{and the links
to }quantum discord, for the experimentally relevant subclass of squeezed-thermal
(STS) states. Moreover, we relate each of these classes to a special
directional QT task, showing that the whole subclass of STS Gaussian
entangled states including those with asymmetric quantum correlation
can be used for QT. By introducing an EPR steering parameter, we establish
an experimental signature to distinguish the states of different classes,
whether EPR steering, entanglement, or discord beyond entanglement.
Finally, we show how one can manipulate the two-mode squeezed EPR
state to cross between the different classes of quantum correlation,
by adding asymmetric amounts of thermal noise to each sub-system. 

Our method connects three Gaussian measures of quantum correlation:
Simon's positive partial transpose (PPT) condition for \emph{entanglement
}\cite{PPT-Simon}, the criterion of Ref. \cite{eprcrit} for \emph{EPR
steering}, and the measure of Giorda and Paris for \emph{discord}
\cite{gausdiscord}. We explain how the PPT condition is equivalent
to a condition on an EPR-type variance. The condition works efficiently
for all Gaussian states due to the introduction of a gain factor $g_{sym}$
which we show gives information about the symmetry of the quantum
correlation, and how the resource can be utilised for QT. Entanglement
can be quantified by the steering measure for each party, and by $g_{sym}$.
Interestingly, we find that {}``quantum A(B) discord'' grows with
better steering from Bob (Alice) to Alice (Bob). We will see that
\textcolor{black}{the steering from $B$ to $A$ and quantum $A$
($B$) discord are asymmetrically sensitive to the thermal noise on
the two systems. }In fact, steering can be created {}``one-way''
using thermal manipulation. We then show that while resources with
symmetic quantum correlation are useful for QT via traditional protocols,
those with asymmetric correlation require asymmetric protocols.

\begin{figure}
\begin{centering}
\includegraphics[width=0.85\columnwidth]{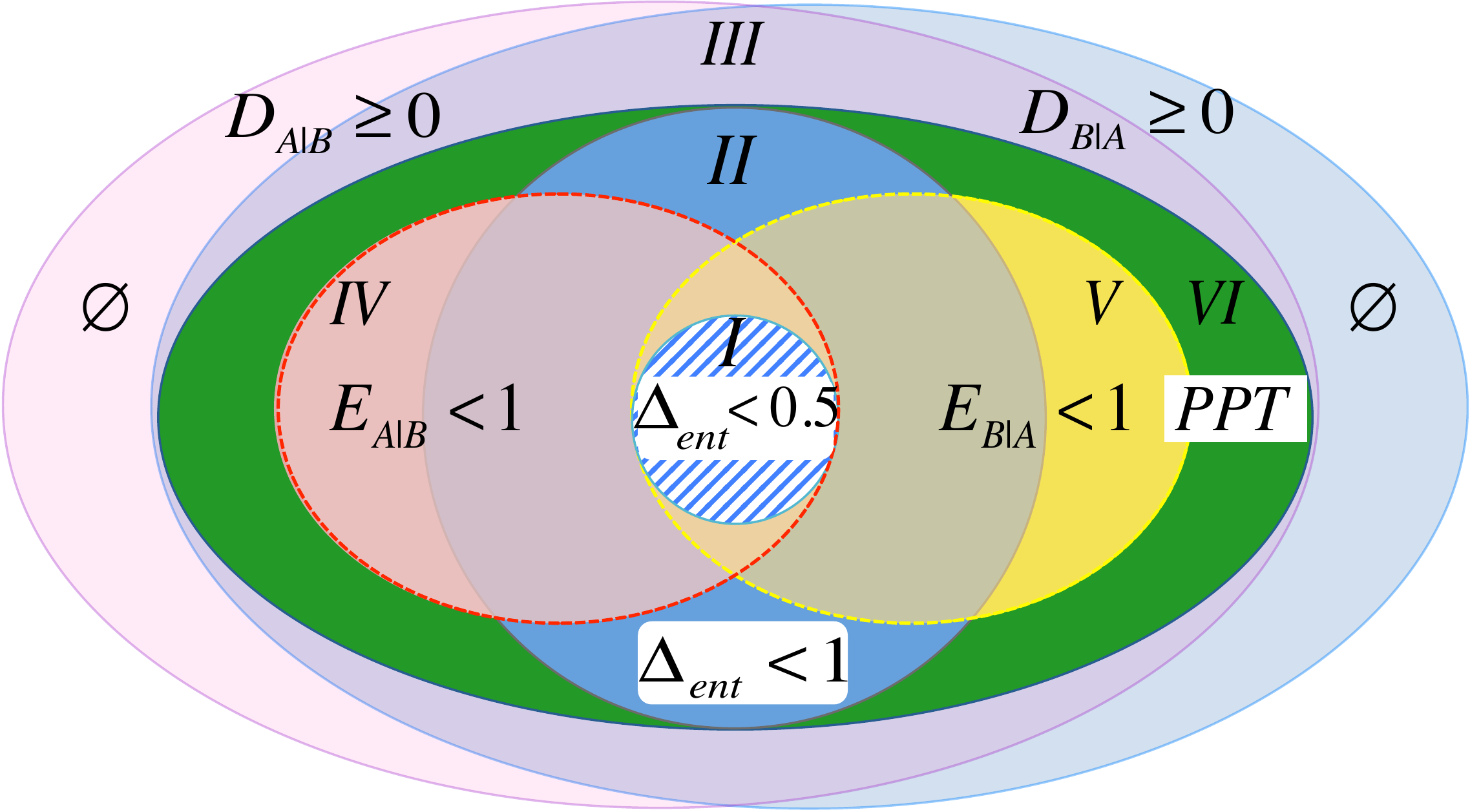}
\par\end{centering}

\begin{centering}

\par\end{centering}

\caption{(Color online) The Venn diagram relations classifying the different
types of quantum correlation for the subclass of Gaussian states.
The larger blue circle $II$ contains states satisfying the Duan criterion
for entanglement $\Delta_{ent}<1$. The inner blue circle $I$ contains
states with the symmetric EPR steering correlation given by $\Delta_{ent}<0.5$.
The set of all entangled states quantified by the PPT criterion $Ent_{PPT}<0$
are contained in the larger green ellipse $VI$. The smaller orange
$IV$ and yellow $V$ ellipses enclose states that display one-way
steering $E_{A|B}<1$ and $E_{B|A}<1$, respectively. Their intersection
(colored yellow) is the set of two-way steerable states, which is
a strict superset of the states in $I$. All two-way steerable states
are a subset of the entangled states quantified by the Duan condition
$\Delta_{ent}<1$. One-way steering states are a strict subset of
the PPT entangled states, and are strictly not contained in the Duan
circle $\Delta_{ent}<1$.\textcolor{red}{{} }\textcolor{black}{The
outer ellipse III contains the set of Gaussian states with non-zero
quantum $A$ and $B$ discord}. All Gaussian states \textcolor{red}{}
except product states are contained in $III$, which is a strict superset
of all Gaussian entangled states \cite{gausdiscord}.\textcolor{red}{{}
} \label{fig:nonloclities-1} }
\end{figure}

All Gaussian properties can be determined from the symplectic form
of the covariance matrix (CM) defined as $C_{ij}=\langle(X_{i}X_{j}+X_{j}X_{i})\rangle/2-\langle X_{i}\rangle\langle X_{j}\rangle$
where $X\equiv(X_{A},P_{A},X_{B},P_{B})$ is the vector of the field
quadratures: 
\begin{equation}
C=\left(\begin{array}{cccc}
n & 0 & c_{1} & 0\\
0 & n & 0 & c_{2}\\
c_{1} & 0 & m & 0\\
0 & c_{2} & 0 & m
\end{array}\right)\label{eq:CM}
\end{equation}
The symplectic invariants are defined by $I_{1}=n^{2}$, $I_{2}=m^{2}$,
$I_{3}=c_{1}c_{2}$, $I_{4}\equiv det(C)=(nm-c_{1}^{2})(nm-c_{2}^{2})$,
and the symplectic eigenvalues $d_{\pm}=\sqrt{\left(\Delta\pm\sqrt{\Delta^{2}-4det(C)}\right)/2}$
with $\Delta=I_{1}+I_{2}+2I_{3}$ $ $\cite{gausdiscord,three criteria Buono}.
Our classification will be exemplified by the Gaussian two-mode squeezed
thermal state (STS) for which $c_{1}=-c_{2}=c$. We thus follow \cite{gausdiscord}
and focus on this subclass of Gaussian states for the remainder of
the paper. The covariance matrix elements in the STS case are $n=(2n_{A}+1)cosh^{2}(r)+(2n_{B}+1)sinh^{2}(r)$,
$m=(2n_{B}+1)cosh^{2}(r)+(2n_{A}+1)sinh^{2}(r)$, $c=(n_{A}+n_{B}+1)sinh(2r)$,
where $n_{A},\ n_{B}$ are the average number of thermal photons for
each system and $r$ denotes the squeezing parameter. Here, we normalise
the vacuum fluctuations so that $\Delta X\Delta P\geq1$. We can specify
Simon's PPT criterion for \emph{entanglement} as \cite{three criteria Buono}
\begin{align}
Ent_{PPT} & =(nm-c^{2})^{2}+1-\left(n^{2}+m^{2}+2c^{2}\right)<0,\label{eq:PPT_ent}
\end{align}
which becomes a necessary and sufficient condition for Gaussian states
\cite{PPT-Simon}. According to the PPT criterion (\ref{eq:PPT_ent}),
a two-mode STS is entangled iff $r$ exceeds the following threshold
value: $cosh^{2}(r_{ent})=\frac{(n_{A}+1)(n_{B}+1)}{n_{A}+n_{B}+1}$
\cite{discordnoise}. The complete set of PPT entangled states is
depicted as contained within the green ellipse of Fig. \ref{fig:nonloclities-1}.
This set is not exhaustive for Gaussian states as seen by the values
for $Ent_{PPT}$ versus the thermal noises $n_{A}$ and $n_{B}$ shown
in Fig. 2a \cite{gausdiscord}. 

Entanglement can also be determined using an EPR-type correlation
\cite{duan,proof for product form}. On defining the weighted difference
variance $\Delta^{2}(X_{A}-gX_{B})=n-2gc+g^{2}m=\Delta^{2}(P_{A}+gP_{B})$,
entanglement between modes $A$ and $B$ is confirmed when\textcolor{red}{}
\begin{eqnarray}
Ent_{g}^{A|B}= & \Delta^{2}(X_{A}-gX_{B})/(1+g^{2})< & 1.\label{eq:Entg}
\end{eqnarray}
Here $g$ is an arbitrary real constant but which can be optimally
chosen to minimise the value of $Ent_{g}^{A|B}$. For the restricted
subclass of Gaussian EPR resources, there is symmetry between the
$X$ and $P$ moments so that a single $g$ suffices. With the optimal
choice of $g=g_{sym}^{A|B}\equiv\left(n-m+\sqrt{(n-m)^{2}+4c^{2}}\right)/2c$,
it is straightforward to show that the entanglement bounds of $Ent_{g}<1$
and $Ent_{PPT}<0$ ($\tilde{d}_{-}=1$, obtained by replacing $I_{3}\rightarrow-I_{3}$
in the formula for $d_{-}$) are equivalent. Note that the entanglement
between modes $A$ and $B$ can be also confirmed when $Ent_{g'}^{B|A}=\Delta^{2}(X_{B}-g'X_{A})/(1+g'^{2})<1,$
which is the same threshold as for $Ent_{g}^{A|B}$ but with $g'=g_{sym}^{B|A}\equiv\left(m-n+\sqrt{(m-n)^{2}+4c^{2}}\right)/2c=1/g_{sym}^{A|B}$.
This is to be expected: Entanglement is by definition a quantity shared
between two systems, and its PPT measure does not account for the
directional properties associated with quantum correlation. 

Where one has complete symmetry between the systems, $n=m$ and $g_{sym}^{A|B}=1$.
The PPT criterion (\ref{eq:Entg}) for entanglement then reduces to
the measure of {}``Duan entanglement'' \cite{duan,three criteria Buono}
\begin{equation}
\Delta_{ent}=\left[\Delta^{2}(X_{A}-X_{B})+\Delta^{2}(P_{A}+P_{B})\right]/4<1.\label{eq:duan}
\end{equation}
\textcolor{black}{Resources with the property (\ref{eq:duan}) are
required for the CV quantum teleportation (QT) of a coherent state,
as achieved using the standard protocol of Braunstein and Kimble \cite{bk}.
The STS squeezing threshold for Duan entanglement is $ $$r>r_{QT,duan}=ln\sqrt{n_{A}+n_{B}+1}$.
These states are depicted as enclosed within the dark blue circle
$II$ of Fig. 1. Sufficiently asymmetric systems (where $n\gg m$)
may arise for example when coupling massive objects to laser pulses,
and may require the full PPT entanglement test (outside the blue circle
$II$, but within the green ellipse) as illustrated in Fig. 1 \cite{thermal epr}.
}
\begin{figure}[t]
\begin{centering}
\includegraphics[width=0.53\columnwidth]{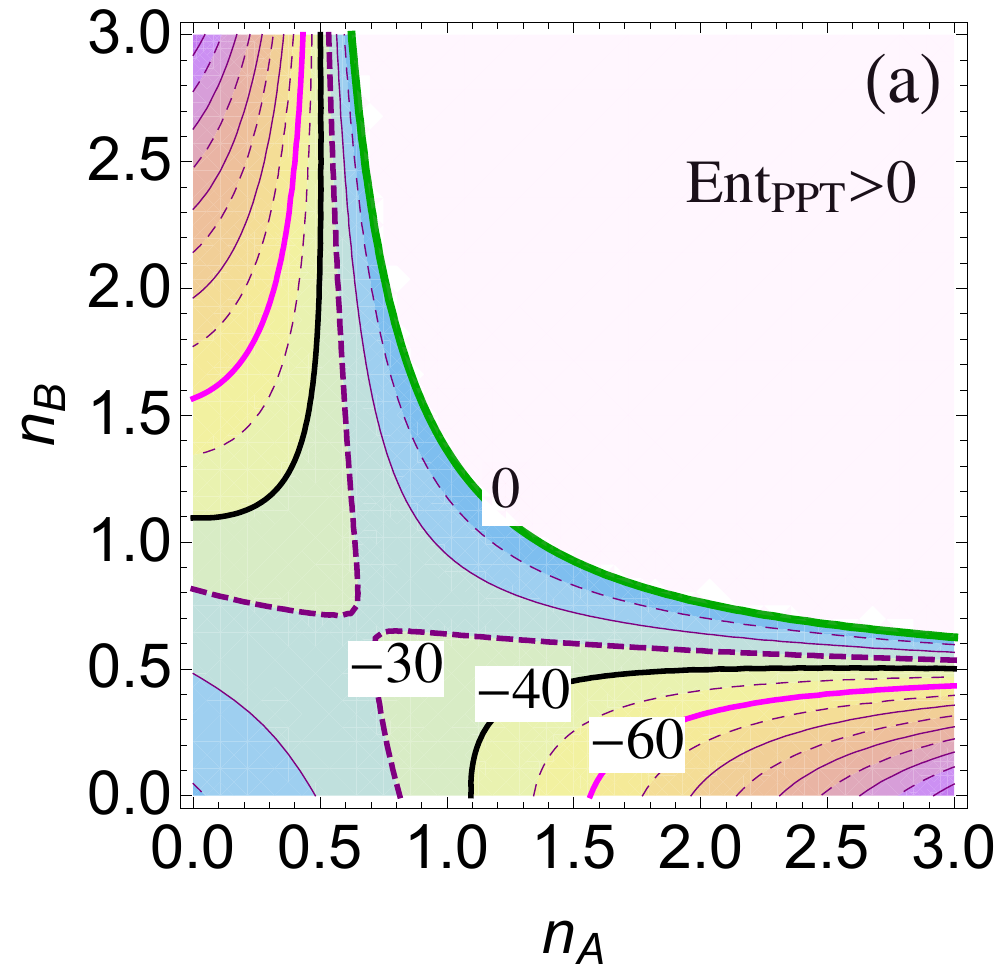}\includegraphics[width=0.53\columnwidth]{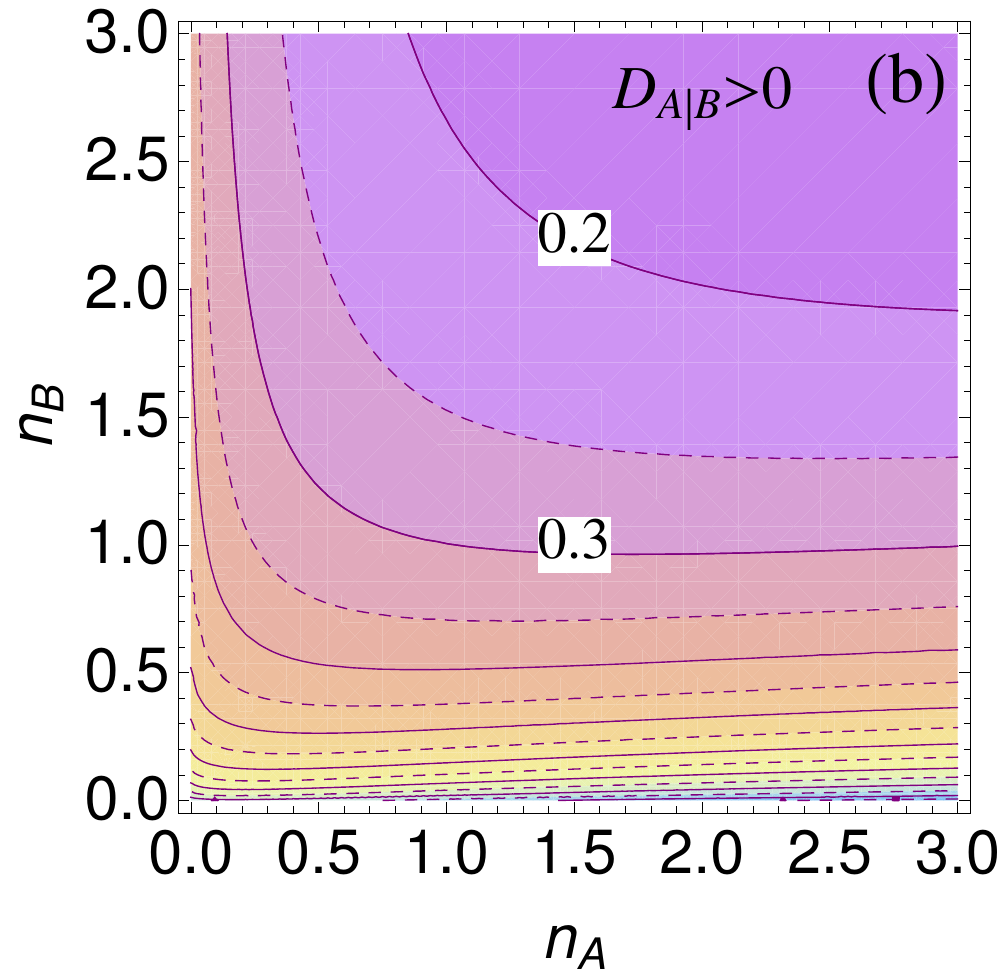}
\par\end{centering}

\begin{centering}
\includegraphics[width=0.53\columnwidth]{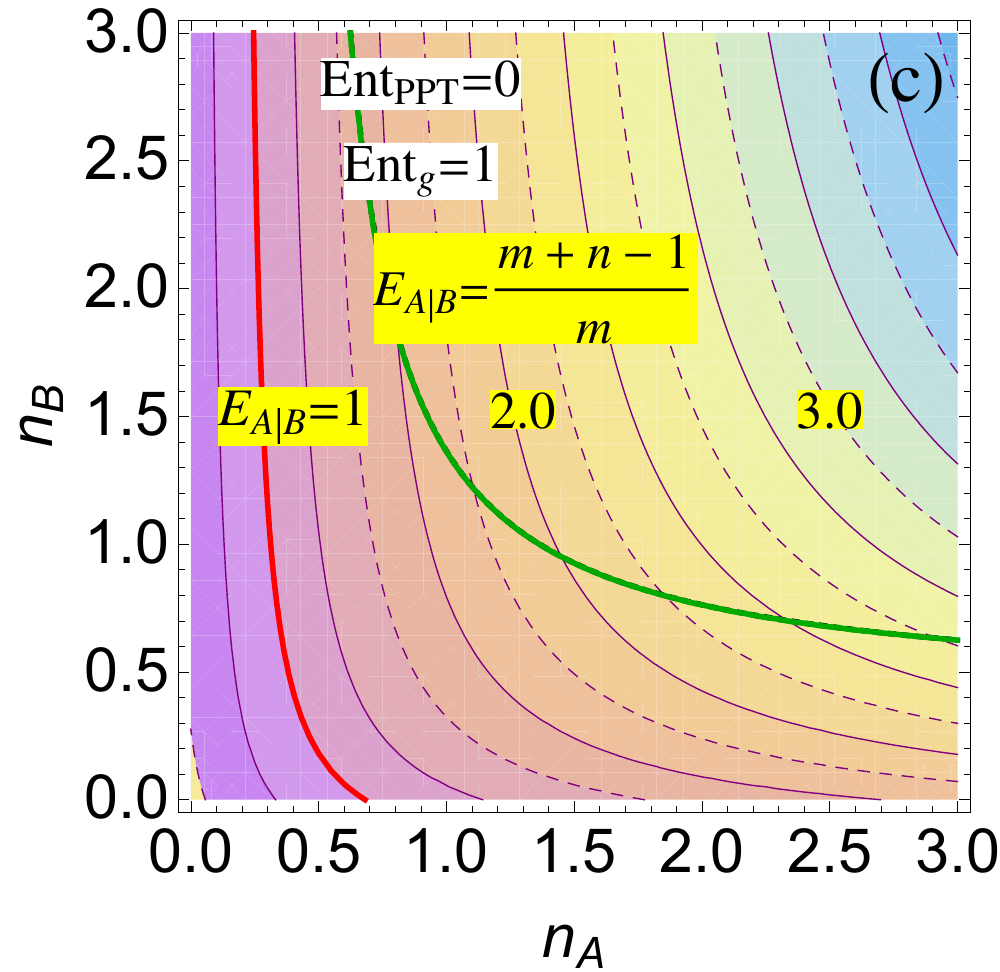}\includegraphics[width=0.53\columnwidth]{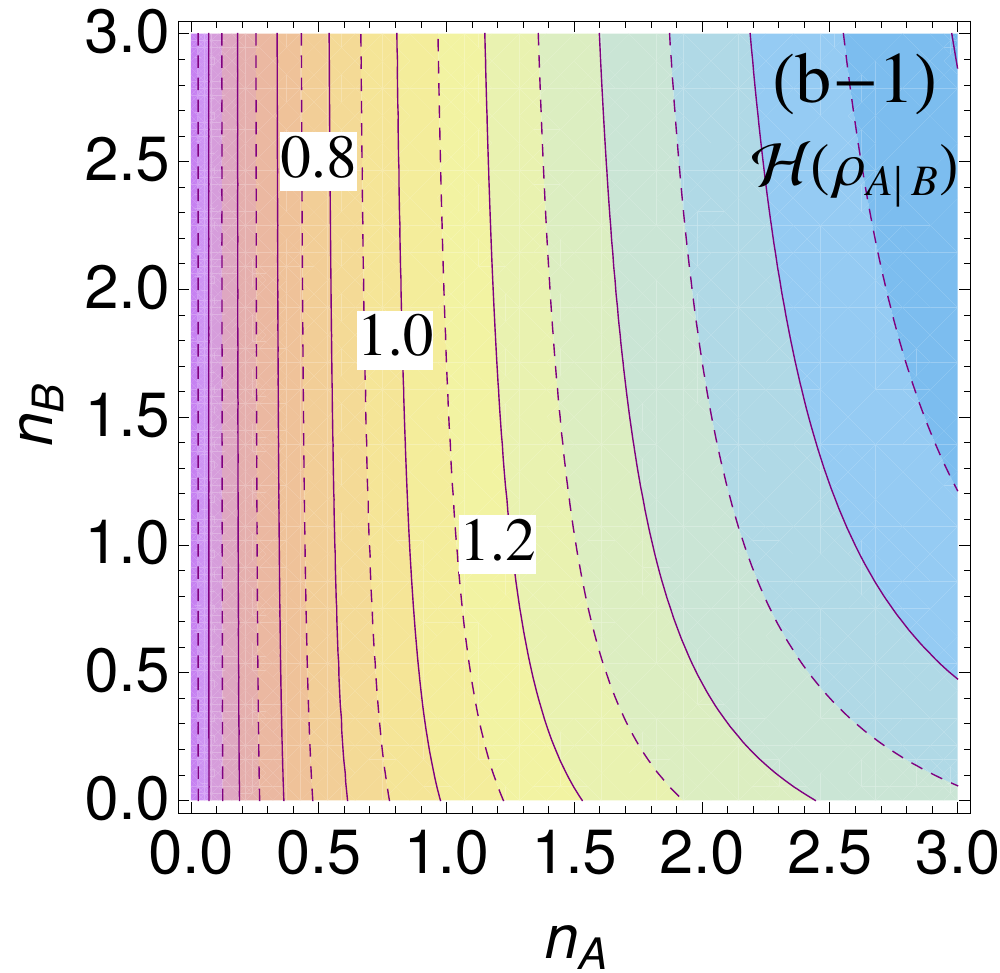}
\par\end{centering}

\begin{centering}
\includegraphics[width=0.52\columnwidth]{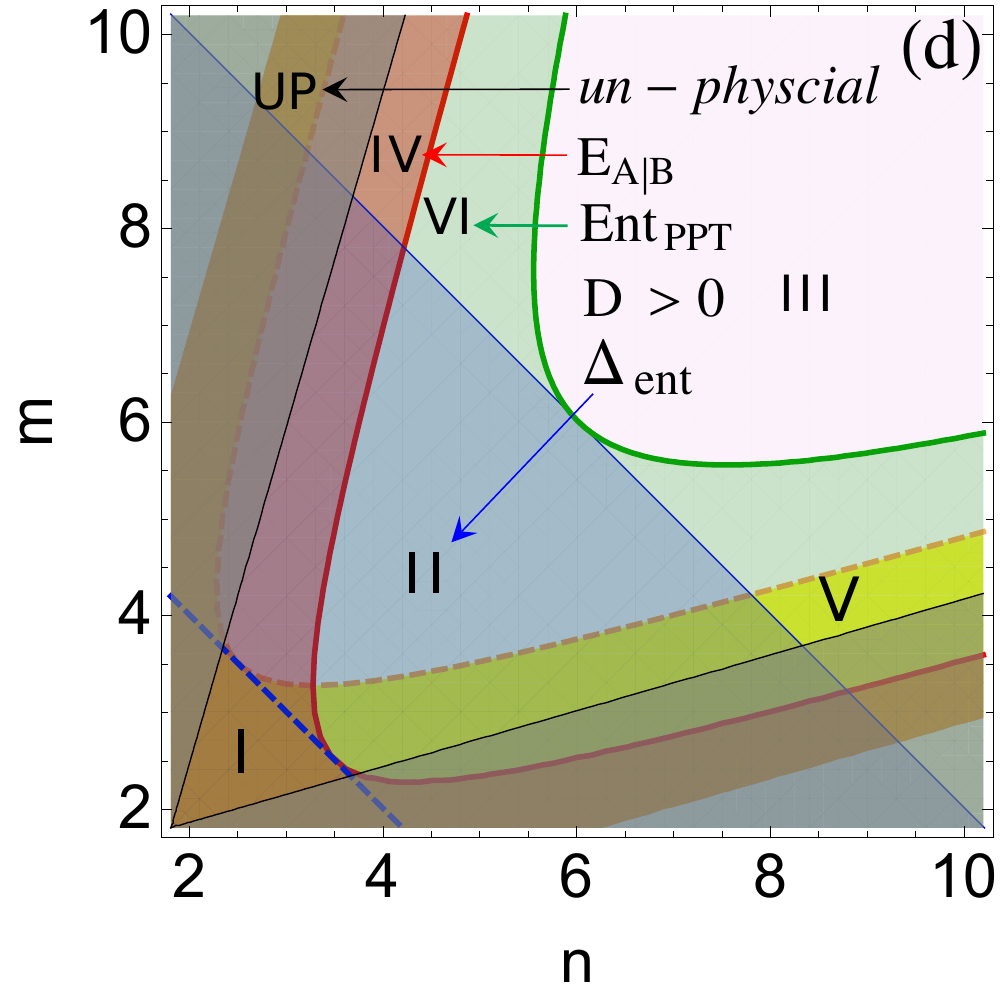}\includegraphics[width=0.53\columnwidth]{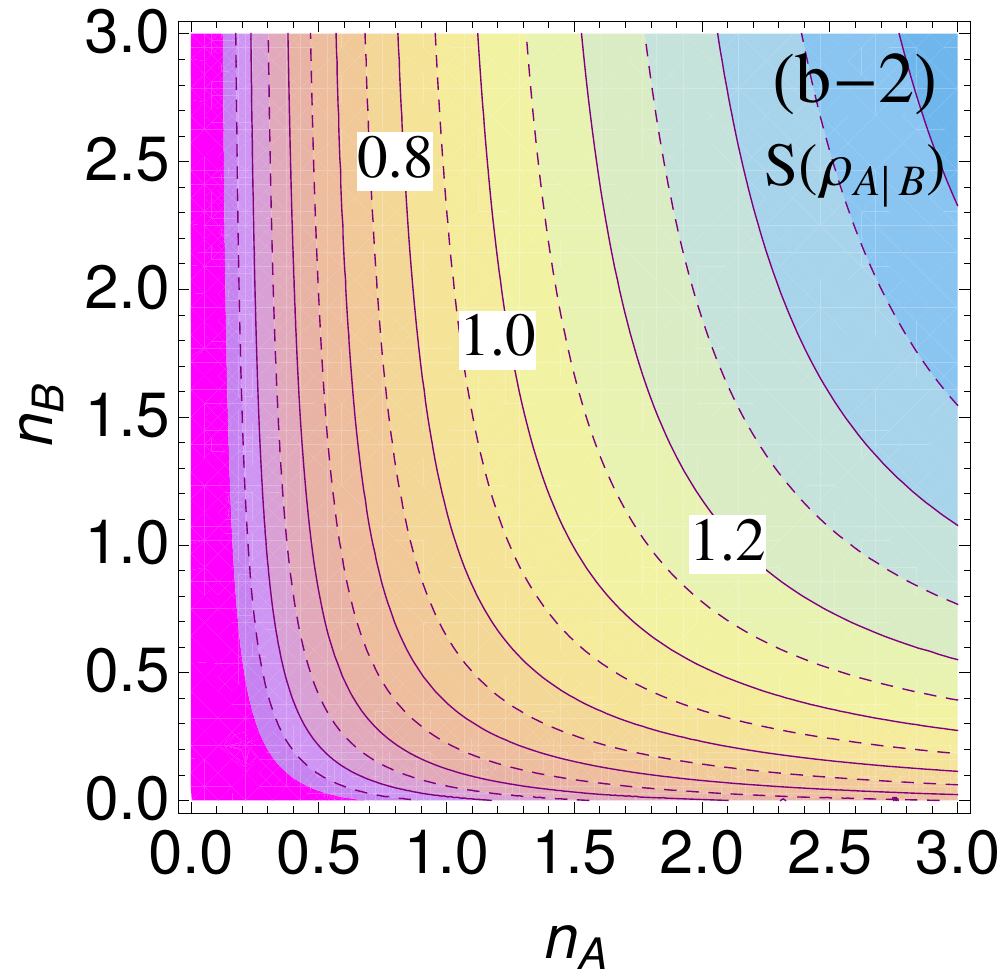}
\par\end{centering}

\caption{(Color online) Contour plots show the effect of asymmetric noises
$n_{A}$ and $n_{B}$ on quantum correlation, for the two-mode STS
with $r=0.6$: (a) entanglement measured by $Ent_{PPT}$, (b) discord
measured by $D_{A|B}$, and (c) the steering parameter $E_{A|B}$.
In (c), the states can be used as quantum resources with EPR steering
(below red curve), entanglement (below green curve), or discord beyond
entanglement (above green curve) as explained in text\textcolor{black}{.
(d) shows the different regions certified by criteria of steering,
entanglement, discord, and unphysical CMs (light gray area $UP$)
}versus $n$ and $m$. The terms contributing to discord, $S(\rho_{A|B})$
and $\mathcal{H}(\rho_{A|B})$, are shown in (b-1) and (b-2), and
are discussed in the Supplemental Materials.\label{fig:two-noises}}
\end{figure}

Quantum discord\emph{ }is by definition a measure of asymmetric quantum
correlation between the two subsystems \cite{discord}. The {}``quantum
A discord'' that considers the conditional information for Alice's
system $A$ based on measurements on system $B$ by Bob, has been
derived for a Gaussian state by Giorda and Paris as \cite{gausdiscord}
\begin{equation}
D_{A|B}=f(m)-f(d_{+})-f(d_{-})+f\left(z\right),\label{eq:discord}
\end{equation}
where $z=\frac{n+mn-c^{2}}{m+1}$ and $f(x)=(\frac{x+1}{2})ln(\frac{x+1}{2})-(\frac{x-1}{2})ln(\frac{x-1}{2})$.
With the exchanging $m\leftrightarrow n$ and hence $I_{1}\leftrightarrow I_{2}$,
we obtain the result for the B discord $D_{B|A}$. Quantum $A$ discord
is obtained for all bipartite Gaussian states that are not product
states, although there are non-entangled states that have nonzero
discord \cite{gausdiscord}. The quantum discord is the difference
between two classically-equivalent definitions of conditional entropy
\cite{discord,discord vedral,gausdiscord}. Denoting the von Neumann
entropy of the quantum state $\rho$ by $S(\rho)$, the first $S(\rho_{A|B})\sim f(d_{+})+f(d_{-})-f(m)$
arises from using the definition of mutual information based on the
bipartite state $\rho_{AB}$. The second arises from quantisation
of the expressions for the conditional entropy: $\mathcal{H}(\rho_{A|B})=\sum_{k}p_{B}(k)S(\rho_{A|k})\sim f(\sqrt{z})$
where $p_{B}(k)$ is the probability of result $k$ for a measurement
at $B$, and $S(\rho_{A|k})=\sum_{i}p(i|k)S(\rho_{i|k})$ where $p(i|k)$
is the conditional probability of outcome $i$ at A given the result
$k$ at B. The discord (\ref{eq:discord}) is obtained by minimising
the mismatch over all Gaussian measurements. The terms in the quantum
$A$ discord $\mathcal{H}$ quantify the available information for
the conditional state of $A$ after measurement on $B$, and also
reflect uncertainty in measurements of Alice when Bob's outcome $k$
is known. 

Interestingly, this reminds us of the other asymmetric nonlocality,
EPR steering\emph{ }from $B$ to $A$ \cite{hw-steering,EPRsteering,epr},
which is realized for Gaussian systems iff \cite{hw-steering,eprcrit}
\begin{equation}
E_{A|B}=\Delta_{inf}X_{A|B}\Delta_{inf}P_{A|B}<1\label{eq:EPR steer}
\end{equation}
Here $\Delta_{inf}^{2}X_{A|B}=\sum_{k}p_{B}(k)\Delta^{2}(X_{A}|k)$
where $\Delta^{2}(X_{A}|k)$ is the variance of the conditional distribution
for Alice's {}``position'' $X_{A}$ conditional on the result $k$.
The measurement at $B$ is selected to minimise the quantity $\Delta_{inf}^{2}X_{A|B}$.
The $\Delta_{inf}^{2}P_{A|B}=\sum_{k'}p_{B}(k')\Delta^{2}(P_{A}|k')$
is defined similarly, for the momentum $P_{A}$. The states with the
property (\ref{eq:EPR steer}) are depicted by the small orange ellipse
of Fig. 1. For Gaussian states, we can write $\Delta_{inf}^{2}X_{A|B}=\Delta^{2}(X_{A}-gX_{B})$
and $\Delta_{inf}^{2}P_{A|B}=\Delta^{2}(P_{A}+gP_{B})$ where $g$
is a real constant \cite{rrmp,eprcrit}, noting that for the restricted
subclass $E_{A|B}(g)=\Delta_{inf}^{2}X_{A|B}=\Delta_{inf}^{2}P_{A|B}=n+g^{2}m-2gc$.
The optimal measurement is defined by the optimal $g$$ $. The quantity
$E_{A|B}(g)$ is minimized to $E_{A|B}=n-c^{2}/m$ by the optimal
factor $g=c/m$ \cite{three criteria Buono,eprcrit}, and its smallness
gives a measure of the degree of nonlocal correlations. Ideally, it
becomes zero in the limit of large $r$. As with discord, we obtain
the result for the steering from $A$ to $B$ by interchanging parameters:
$E_{B|A}(g')=\Delta_{inf}X_{B|A}\Delta_{inf}P_{B|A}=m-c^{2}/n$ where
$g'=c/n$ (small yellow ellipse of Fig. 1).

Two-way steering is confirmed when both $E_{A|B}(g_{s})<1$ and $E_{B|A}(g_{s}')<1$,
given by the yellow intersection of the two smaller ellipses of Fig.
1. For $g_{s}=1$, we have $E_{A|B}=E_{B|A}=n+m-2c=2\Delta_{ent}$
and hence $\Delta_{ent}<0.5$ is a criterion sufficient to confirm
two-way EPR steering. This is also the Grosshans and Grangier condition
required of an EPR resource for the secure teleportation (ST) of a
coherent state \cite{Grangier clone-1}. In that case, a teleportation
fidelity $F=1/\left(1+\Delta_{ent}\right)>2/3$, as opposed to $F>1/2$
for QT, is needed.\textcolor{black}{{} To satisfy }$E_{A|B}(g)<1$ or
$E_{B|A}(g)<1$ requires t\textcolor{black}{he squeezing $r$ to exceed
the threshold value given by $r_{A|B}$ and $r_{B|A}$ respectively,}
where $cosh^{2}\left(r_{A|B}\right)=\frac{(2n_{A}+1)(n_{B}+1)}{1+n_{B}+n{}_{A}}$
or $cosh^{2}\left(r_{B|A}\right)=\frac{(n_{A}+1)(2n_{B}+1)}{1+n_{B}+n{}_{A}}$.
The two-mode STS \textcolor{black}{with} $ $$r>\{r_{A|B},r_{B|A}\}_{max}$
can be used to produce two-way steering, which is only possible for
sufficient symmetry given by $|n_{A}-n_{B}|<1/2$. The states satisfying
the strongly symmetric EPR correlation $\Delta_{ent}<0.5$ are depicted
by the centre light blue circle $I$ of Fig. 1. This requires t\textcolor{black}{he
squeezing $r$ to exceed the threshold value} $ $$r>r_{ST,duan}=ln\sqrt{2(n_{A}+n_{B}+1)}$,
and $\Delta_{ent}<0.5$ is not therefore a necessary condition for
two-way steering. Two-way steering is possible when $\{r_{A|B},r_{B|A}\}_{max}<r<r_{ST,duan}$,
as shown by the yellow region not contained in $I$ (Fig. \ref{fig:nonloclities-1}).\textcolor{red}{}

We note that the inequality (\ref{eq:Entg}) to determine Gaussian
entanglement with $g=g_{sym}^{A|B}$ and inequality (\ref{eq:PPT_ent})
both require $nm-c^{2}+1-n-m<0$. This can be written as a bound on
the steering parameter: 
\begin{equation}
E_{A|B}\equiv E_{A|B}(g)<\frac{m+n-1}{m},\label{eq:ent_E}
\end{equation}
with factor $g=c/m$. This can be also written as $E_{B|A}\equiv E_{B|A}(g')<\left(m+n-1\right)/n$
with the optimal gain factor $g'=c/n$. Hence, we establish a unified
experimental measure of quantum correlation: EPR steering if $E_{A|B}<1$
is satisfied (below the red curve in Fig. \ref{fig:two-noises}c);
entanglement if $E_{A|B}<\left(m+n-1\right)/m$ (below the green curve
in Fig. \ref{fig:two-noises}c) and discord beyond entanglement if
$E_{A|B}>\left(m+n-1\right)/m$ (above the green curve in Fig. \ref{fig:two-noises}c).

Pure states or symmetric states ($n=m$) imply $g_{sym}=1$, so that
criteria (\ref{eq:PPT_ent}), (\ref{eq:Entg}), (\ref{eq:ent_E})
with $g=tanh(2r)$ are all reduced to the Duan entanglement criterion
$\Delta_{ent}<1$. All such {}``Duan-entangled'' states are contained
within the blue circle $II$. These states are required for traditional
CV QT \cite{bk,three criteria Buono}. The directional correlation
happens for asymmetric mixed states, which create the ellipses of
Fig. 1 outside the blue circle $II$. 

We now present the conditions for creating the states of each class
of quantum correlation. The effect of thermal noises $n_{A}$, $n_{B}$
on entanglement, discord and steering for the STS with $r=0.6$ is
illustrated in Fig. \ref{fig:two-noises}. Generally, the presence
of asymmetric noises creates the possibility of asymmetric steering/
discord, making steering/ disturbance from A to B more difficult than
that from $B$ to $A$.  Entanglement is absent for $Ent_{PPT}\geq0$,
the region above the green curve in Fig. 2a. All regions show {}``quantum
A discord'', given by $D_{A|B}>0$ (Fig. 2b) \cite{gausdiscord}.
Thermal noises tend to suppress entanglement, for which the dependence
on $n_{A}$ and $n_{B}$ is symmetric. However, the effect on the
discord is more complex and asymmetrical. \textcolor{black}{We can
see that $D_{A|B}$ is maximised when most of thermal noise is placed
on the unmeasured system A.}

\textcolor{black}{Figure \ref{fig:two-noises} (c) shows the behaviour
of the steering parameter }$E_{A|B}$\textcolor{black}{}. T\textcolor{black}{he
sensitivity to the noises is asymmetrical and {}``one-way steering''
(the states contained in the smallest left ellipse of Fig. 1 but exclusive
of the right one) is evident. }The value of $E_{A|B}$ is minimised
(and steering increased) when\textcolor{black}{{} most of thermal noise
is placed on the system $B$}, since $E_{A|B}<\frac{m+n-1}{m}\sim1$
when\textcolor{black}{{} }$m\gg n$\textcolor{black}{.}

The behavior of discord is strongly related to steering (Fig. \ref{fig:two-noises}).
We note the similarity between the conditional entropy $\mathcal{H}(\rho_{A|B})$
(Fig. \ref{fig:two-noises}(b-1)) and $E_{A|B}$ (Fig. \ref{fig:two-noises}c).
As steering increases (so that $E_{A|B}\rightarrow0$) the variances
of the conditional distribution are reduced \cite{eprcrit,rrmp}.
We find that for better steering of Alice by Bob, quantum discord
becomes larger (see Supplemental Materials \cite{proof1}). This is
consistent with the picture that more of the EPR-type disturbances
happen to Alice's system because of Bob's measurements.  

\begin{figure}[t]
\begin{centering}
\includegraphics[width=0.85\columnwidth]{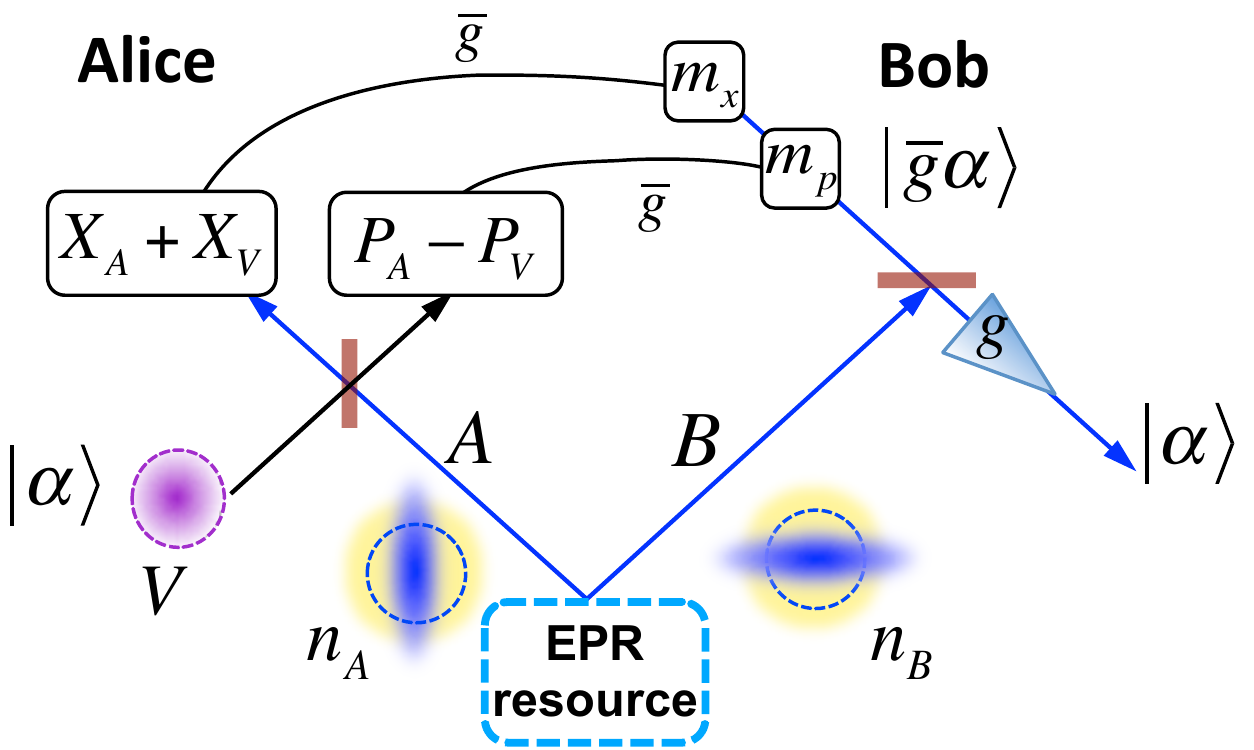}
\par\end{centering}

\caption{Scheme for quantum teleportation with direction distinguished. Here
we depict CV quantum teleportation of Victor's coherent state $|\alpha\rangle$
at $V$ to an amplified coherent state $|\bar{g}\alpha\rangle$ at
Bob's location, which can then be attenuated by factor $g=1/\bar{g}$.
The Bell measurement is defined at $A$ as in Ref. \cite{bk}. \textcolor{red}{}The
scheme uses a resource with the directional entanglement specified
by $g_{sym}^{B|A}=\bar{g}\geq1$. \textcolor{red}{}The resource is
generated by adding asymmetric noise ($n_{B}>n_{A})$ to the output
channels of the EPR source. The maximum fidelity of the scheme is
$F_{g}=1/\bar{g}^{2}$ and is achieved when $E_{B|A}(\bar{g})=\bar{g}^{2}-1$;
QT is achieved for $F>0.5$ which is satisfied iff $Ent_{PPT}<0$.
 \label{fig:QT}}
\end{figure}

Finally, we emphasize potential applications of asymmetric correlation.
We show in the Supplementary Materials \cite{proof1} that the directional
entangled states\textcolor{red}{{} }are useful as a resource for the
quantum teleportation of a coherent state from Alice to Bob (if $g_{sym}^{A|B}$$\leq1$),
or from Bob to Alice (if $g_{sym}^{A|B}\geq1$). This is achieved
using the asymmetric protocol of Fig. 3. We leave open the question
of whether the asymmetric value of discord may also produce directional
quantum tasks only successful either from Alice to Bob, or Bob to
Alice. 

In conclusion, we have established classes of CV Gaussian quantum
correlation, determined how to signify and generate states of a given
class, and shown how the states of each entanglement class can be
utilised for a quantum teleportation task. We explored the relation
between two asymmetric nonclassical correlations, steering and discord.
Our results suggest asymmetric correlations such as EPR steering and
discord to be promising candidates for quantum tasks requiring a directional
operation.\textcolor{green}{}
\begin{acknowledgments}
We thank the Australian Research Council for funding via Discovery
and DECRA grants. Q. Y. H thanks support from the National Natural
Science Foundation of China under grants No.11121091 No.11274025. \end{acknowledgments}

\end{document}